**Revisiting Classical Two-phase and Kerner Three-phase Traffic Flow Theories: A Comparison of Pre-crash and Normal Traffic Conditions**

**Md Mahmud Hossain**
(*Corresponding Author*)
Texas A&M Transportation Institute
Email: m-hossain@tti.tamu.edu
ORCID: 0000-0002-2737-6951

**Kazi Tahsin Huda**
Department of Engineering Technology and Construction Management, University of North Carolina at Charlotte, USA
Email: khuda@uncc.edu

**Moinul Hossain**
Department of Civil and Environmental Engineering, Islamic University of Technology, Bangladesh
Email: moinul@iut-dhaka.edu
ORCID: 0000-0002-1178-9516

**Yasunori Muromachi**
Department of Built Environment, School of Engineering, Tokyo Institute of Technology
E-mail: ymuro@enveng.titech.ac.jp

# Revisiting Classical Two-phase and Kerner Three-phase Traffic Flow Theories: A Comparison of Pre-crash and Normal Traffic Conditions

**ABSTRACT**

Extensive research has been conducted to develop statistical and artificial intelligence-based models for predicting short-term crash probabilities using fundamental traffic flow variables and their associated descriptive statistics and mathematical transformations. However, there has been limited exploration into whether and how the fundamental relationships within traffic flow theories vary between pre-crash and normal traffic conditions. This study reevaluates four classical two-phase traffic flow theories and employs two methods from Kerner's three-phase traffic flow theory to compare their characteristics in the context of pre-crash and normal traffic conditions. The investigation is centered around the Shibuya 3 and Shinjuku 4 routes within the Tokyo Metropolitan Expressway. Data from both crashes and detectors was collected over a six-month period, spanning from March 2014 to August 2014. The findings reveal that data from the nearest downstream detectors to the crash locations provided a superior fit for pre-crash data compared to the upstream detectors. Notably, a noticeable decrease in goodness-of-fit was observed when compared to normal traffic conditions. In pre-crash scenarios, wide-moving jams exhibited a faster propagation from downstream to upstream, distinct from the patterns observed in normal traffic conditions. Furthermore, pre-crash data displayed a higher standard deviation in the calculated wide-moving jam velocities compared to normal traffic conditions. These insights have the potential to be highly valuable in the development of predictive crash models and in the estimation of traffic volumes on freeways across various timeframes, accounting for both free-flow and congested traffic scenarios.

# 1 INTRODUCTION

## 1.1 Brief description of traffic flow theories

Traffic flow theories are indispensable parts of the foundation of traffic science, serving as the backbone of transportation engineering. They have various applications in systems where numerous vehicles interact with one another [1]. These theories distinguish between microscopic and macroscopic models and their associated variables. Macroscopic models provide an overview of the entire traffic stream and elucidate how the parameters (such as speed, flow, and density) relate to one another [2]. On the other hand, microscopic models delve into the behavior of individual vehicles within a road network and their interactions [3]. These theories are inevitable for understanding the fundamental characteristics of roadways. More than 80 years ago, the first macroscopic traffic flow theory, as proposed by Greenshields et al. [4], assumed a linear relationship between speed and density. Greenberg [5] suggested a simple logarithmic relationship, but it didn't perform well in predicting speed at low-density conditions. Underwood [6] posited an exponential relationship, but it had limitations in high-density traffic scenarios. Subsequently, Drake et al. [7] associated traffic flow models with a bell-shaped normal distribution curve. Over time, more complex models have been developed, yet these four models remain widely used by researchers and practitioners. All these models fundamentally describe two distinct traffic phases: the 'free phase' where drivers can travel at their preferred speeds comfortably, and the 'jam phase' which characterizes states where vehicles move at relatively low speeds or come to a standstill.

The three-phase traffic flow theory offers a more realistic and accurate portrayal of traffic flow bason upon real-time traffic observations [8]. Rehborn et al. [9] employed four methods- detector based, correlation, flow-density and graphical method- for identifying the wide moving jam velocity of traffic flow in Kerner's three-phase traffic flow theory. Conventional two-phase traffic theories fall short in explaining the complex phase transitions and jam phenomena observed in real traffic [10]. Kerner's three-phase traffic flow theory addresses these limitations and rectifies them [11]. Drawing from empirical data, this theory categorizes traffic flow into three distinct phases: free flow, synchronized flow, and wide moving jam. The free flow phase occurs when a driver can travel at a speed comfortable to them, considering factors like speed limits, weather conditions, vehicle type, passenger count, and engine capacity. Congested traffic is indicated by vehicle speeds lower than the slowest vehicle in the free flow state [12]. A wide moving jam originates from bottlenecks and maintains a mean velocity as it moves upstream, even amidst other traffic states or freeway bottlenecks [9]. Synchronized flow, on the other hand, is typically fixed at the bottleneck and does not exhibit wide moving jam characteristics [13]. Practical online applications like ASDA/FOTO have been developed based on these observations to monitor and identify different traffic flow phases [14].

## 1.2 Applications of traffic flow theories

In addition to their fundamental role in traffic flow theories, the core variables of traffic flow such as speed, flow, and density hold significant relevance in traffic engineering, particularly concerning road safety. Researchers are actively exploring the use of these traffic flow variables to predict road crashes in real-time, a concept referred to as real-time crash prediction models. Certain combinations of speed, flow, and density can lead to uncomfortable driving conditions, potentially resulting in errors and, ultimately, collisions [15,16]. In developed regions, numerous access-controlled roads like freeways and expressways are equipped with a substantial array of sensors and cameras, providing real-time traffic flow variable data that can serve as input for these models [17]. Research in real-time crash prediction has primarily focused on distinguishing pre-crash data from normal traffic data using various prediction models [18–24]. This research also involves identifying different types of crashes [25–27], understanding the mechanics of crashes [19,28–30], and exploring potential countermeasures to mitigate hazardous traffic conditions [20,31]. The process of distinguishing hazardous traffic conditions from normal ones can be broadly categorized into statistical approaches, including nonparametric Bayesian statistics [32], aggregate log-linear models [28], generalized estimating equations [21], matched case-control logistic regression [29,33], seemingly unrelated negative binomial regression [22], and artificial intelligence-based methods, such as probabilistic neural networks [32,34], normalized radial basis functions [23], naive Bayes [30], Bayesian Networks [35], among others.

Regarding studies aimed at understanding the mechanics of crashes, Hossain and Muramachi [36] employed Classification and Regression Trees to identify clusters of hazardous traffic conditions. Abdel-Aty et al. [37] discovered specific high-speed and low-speed conditions closely linked to collisions. Pande and Abdel-Aty [27] and Lee et al. [29] determined that crashes related to lane changes are significantly influenced by average speeds upstream and downstream, along with variations in occupancy on adjacent lanes and the standard deviation of volume and speed at a location downstream where the crash occurs. These investigations collectively provide substantial evidence of the substantial differences between pre-crash and normal traffic conditions. Such efforts are underway to implement real-time interventions to restore traffic to normal conditions. However, due to the associated high risks, such studies cannot be conducted using real-world data. As a solution, researchers have found it feasible to employ micro-simulation to simulate pre-crash scenarios and then apply various interventions. These interventions include variable speed limits [20,22,28,34], ramp metering [29,31], variable message signs [20], and more, with the aim of restoring traffic conditions to a normal state.

*1.3 Study objectives*

Based on the earlier discussion, there is a pressing need for a more profound comprehension of pre-crash traffic conditions through the lens of fundamental traffic flow theories. Surprisingly, there has been no previous attempt to elucidate the distinctions between pre-crash and normal traffic conditions from the perspective of these fundamental traffic flow theories. The primary goal of this research is to revisit traffic flow theories for both pre-crash and normal traffic scenarios. The specific objectives encompass:

- Developing and comparing classical two-phase traffic flow relationships for data pertaining to pre-crash and normal traffic conditions. This involves deriving four classical traffic flow theories (Greenshields, Greenberg, Underwood, and Bell-shaped models) individually for pre-crash and normal traffic data.
- Assessing the longitudinal variations in traffic flow relationships concerning both temporal and spatial scales for pre-crash traffic data.
- Employing hypothesis tests to detect disparities between pre-crash and normal traffic conditions and discussing any variations that may emerge.
- Applying the same methodology to explore Kerner's three-phase traffic flow theory (detector and correlation-based methods).

Utilizing the normal traffic flow conditions across different timeframes and situations and comparing them with pre-crash conditions to assess the likelihood of crash occurrence is a pivotal concept in the development of road safety. For instance, in crash prediction models, such traffic data is utilized to recognize common pre-crash conditions that can be instrumental in predicting the likelihood of an accident. Therefore, the outcomes of this research have the potential to not only identify the probability of a crash but also facilitate the adoption of preventive measures.

## 2 METHODOLOGY

### 2.1 Deriving linear forms for classical two-phase traffic flow theories

The fundamental relationship among the three primary variables in traffic flow theory: flow (q), speed ($V$) and density ($K$) can be expressed as follows:

$$q = V \times K \tag{1}$$

**Table 1** presents the fundamental non-linear relationships of four classic traffic flow theories, Greenberg, Underwood, and Bell-Shape, with respect to density and speed. The table also includes their linear transformations, where $K_j$ = jam density; $V_f$ = free flow speed; $K_0$ = density at maximum flow; $V_0$ = speed at maximum flow level; $K$ = density and $V$ = speed. It is important to mention that this study employs occupancy data in place of density, and the notations $K_j$, $K_0$, and $K$ refer to occupancy during jam density, density at maximum flow conditions, and occupancy at all conditions, respectively.

**Table 1.** Transformation of speed-density functions

| Model | Speed-density Function | Transformation |
|---|---|---|
| Greenshields | $K = K_j(1 - \frac{V}{V_f})$ | $K = K_j - K_j \frac{V}{V_f}$ |
| Greenberg | $K = K_j e^{\frac{-V}{V_0}}$ | $lnK = ln(K_j) + \left(-\frac{1}{V_0}\right)V$ |
| Underwood | $K = K_0 ln(\frac{V_f}{V})$ | $K = K_0 ln(V_f) - K_0 ln(V)$ |
| Bell Shape | $K = K_0(2ln\frac{V_f}{V})^{\frac{1}{2}}$ | $K^2 = 2K_0^2 ln(V_f) - 2K_0^2 ln(V)$ |

### 2.2 Methods for Kerner's three-phase traffic flow theory

There exist four widely used methods to determine the wide moving jam velocity within a traffic flow [9]. The flow-density method is particularly well-suited for extended freeway segments in absence of on or off ramps. Meanwhile, the graphical method involves the processing of images using specialized software known as ASDA/FOTO. This study conducted in an urban expressway densely packed with ramps; therefore, only the detector-based and correlation methods were employed. The specific details and explanations of these methods are provided below.

Assume, two detectors *b* and *a* are positioned at sufficiently large distances from each other. The velocity of the downstream jam front can be determined by dividing the distance between these two detectors by the time difference in the registration of the jam front at each detector [38].

$$v_{gr} = \frac{Loc(b) - Loc(a)}{T_a(when\ v_{a(t)} > 30\ kmph) - T_b(when\ v_{b(t)} > 30\ kmph)} \tag{2}$$

Here *Loc(b)* and *Loc(a)* represent the precise positions of the selected detectors, while $T_b$ and $T_a$ denote the specific times, measured in minutes, at which the real-time speeds exceed 30 km/h immediately after a wide moving jam. The velocity of a downstream detector

$b$ can be correlated with its upstream counterpart $a$ as follows [39]:

$$\text{correl}(q)_k^{(b,a)} = \frac{\frac{1}{n}\sum_{i=1}^{n}((q_i^a - \bar{q}^a) \times (q_{i+k}^b - \bar{q}^b))}{\sqrt{(\frac{1}{n}\sum_{i=1}^{n}(q_i^a - \bar{q}^a)^2)} \times \sqrt{(\frac{1}{n}\sum_{i=1}^{n}(q_{i+k}^b - \bar{q}_k^b)^2)}} \tag{3}$$

Here,
$q_i^b$ = Flow of the downstream detector $b$
$\bar{q}^b$ = Average flow of the downstream detector $b$
$q_{i+k}^a$ = Flow of the upstream detector a with a time lag of $k$ minutes
$\bar{q}^a$ = Average flow of the upstream detector $a$
$\bar{q}_k^a = \frac{1}{n}\sum_{i=1}^{n} q_{i+k}^a$ = Average flow of the upstream detector a for the time lag of $k$ minutes ($i$ = Time index, $k$ = Time lag in minutes)

Subsequently, a cross-correlation function can determine the wave velocity for various time lag intervals. The velocity with highest correlation is the downstream jam velocity for the traffic flow.

$$u = 60(\frac{Loc(b) - Loc(a)}{k}) \tag{4}$$

*2.3 Comparing linear regression models of two independent samples*

In statistics, it is a frequent practice to compare the slopes of simple linear regression models. This is often done to assess whether the linear regression lines generated from two different samples are representative of the same population. Assume that two simple linear regression lines, characterized by slopes $b_1$ and $b_2$, have been constructed using two samples following the pattern described in Equation 5.

$$y = \beta x + c \tag{5}$$

Here, $y$ = dependent variable, $\beta$ = slope, $x$ = independent variable, $c$ = intercept. To test if they come from the same population:

- Test the null hypothesis: $H_0$: $= \beta_1 = \beta_2$, i.e., $\beta_1 - \beta_2 = 0$.
- The t-test statistics can be:
$$t = \frac{b_1 - b_2}{\sqrt{s_{b_1}^2 + s_{b_2}^2}} \sim T(n_1 + n_2 - 4) \tag{6}$$
- If the null hypothesis is true then the difference in the slopes will form a normal distribution with mean to be zero, i.e.,
$$\beta_1 - \beta_2 \sim N(0, s_{b_1 - b_2})$$
where,
$$s_{b_1 - b_2} = \sqrt{s_{b_1}^2 + s_{b_2}^2} \tag{7}$$

If p-value is less than a significance level ($\alpha$), it indicates a significant difference in slope. Conversely, if the p-value is equal to or greater than $\alpha$, the null hypothesis cannot be rejected.

**3 DATA**

*3.1 Study location*

The research focuses on the study area encompassing the Shibuya 3 and Shinjuku 4 routes located within the Tokyo Metropolitan Expressway. These routes are centrally located within the Tokyo Metropolitan area, connecting and traversing crucial business and residential areas. Shibuya 3 spans a length of 11.9 kilometers, while Shinjuku 4 covers 13.5 kilometers. Within this study area, there are a total of 210 detectors, most of which are uniformly spaced, with an average inter-detector gap of approximately 250 meters. These detectors collect data related to speed, traffic flow, and occupancy. The dataset for crashes includes details such as date, time, location, vehicles involved, and lane number. Both the crash and detector data were gathered over a period of six months, from March 2014 to August 2014, with a total of 620 crash samples having complete information. Notably, the selected routes are highly controlled in terms of access, ensuring uninterrupted traffic flow. This feature is essential for the development of classical two-phase traffic flow theories.

*3.2 Variable selection*

The general methodology employed in this study encompasses the defining and extraction of data pertaining to 'pre-crash' and 'normal' traffic conditions, followed by the reconstruction of classical two-phase and three-phase traffic flow models for subsequent comparison. In the context of the two-phase theory, the "speed-density" relationship was transformed into a linear form, and hypothesis tests were conducted to assess the significance of differences in slopes between 'pre-crash' and 'normal' traffic conditions. Finally, the differences in characteristics of the curves were explained based on traffic flow parameters. Traffic flow theories rely on input variables such as speed, flow, and density. In this study, even though speed and flow values were obtained, the detectors provided occupancy values in lieu of density. Existing literature suggests that when the proportion of heavy vehicles is low (around 8% in the study area), and the detector length is roughly consistent (which is the case in the study area), density and occupancy can be assumed to exhibit a linear relationship [40]. Therefore, the research team considered occupancy values instead of density when reconstructing the models. For the three-phase theory, the speeds of the wide moving jam propagating upstream from downstream were calculated for both pre-crash and normal traffic conditions in each crashes, and their statistics were compared.

*3.3 Data pre-processing for classical two-phase traffic flow theories*

The authorities provided raw detector data for each lane, which included information on speed in km/hr, flow in terms of the number of vehicles, occupancy as a percentage. This data was recorded and stored in separate files for each day. To prepare the data for analysis, it was necessary to aggregate the data from two lanes for each detector location. This involved calculating the average speed and occupancy of the two detectors and adding the flow values for the two lanes. Additionally, the original dataset needed to be transposed and reorganized for further analysis. Initially, the dataset had data from all detectors for a specific one-minute interval condensed into a single record. To reformat the data, a program was developed in C++ to extract and compile it into a single file, under the following headers: Route number, direction (inbound or outbound), detector ID, date, time, flow, speed, and occupancy. Finally, both the detector data and crash data were imported into the PostgreSQL relational database system. Subsequently, 'pre-crash' and 'normal' traffic condition datasets were generated by running a series of sub-queries to align with their respective definitions.

The 'pre-crash' data were defined as crash data within the vicinity of the crash location shortly before its occurrence time where vicinity refers to a zone bordered by the nearest upstream (UD) and the nearest downstream (DD) detector locations with respect to crash location, as presented with **Figure 1**. Within the study area, there are two lanes in each direction, with one detector installed in each lane, resulting in two detectors at each detector location. Hence, the values for speed and occupancy were averaged, and flow data were summed to derive the final values for speed, flow, and occupancy at each detector location. For each crash incident, pre-crash data was extracted for speed, flow, and occupancy up to the ninth minute prior to the crash. This data is represented as $XD_tY$, where X can be either U (Upstream) or D (Downstream), Y can be S (Speed), F (Flow), or O (Occupancy), and t ranges from 1 to 9. For example, $UD_6S$ represents the speed at the nearest upstream detector 6 minutes before the crash. Regarding the 'pre-crash' data, 18 datasets (2 for upstream and downstream X 9 for the 9 time slices) were created, with each containing 620 records. These datasets were named as $X_t$, where X could be U (Upstream) or D (Downstream), and t ranged from 1 to 9. For instance, the dataset $D_7$ contains speed, flow, and occupancy data for all 620 crashes collected from the nearest downstream detector seven minutes before the crash.

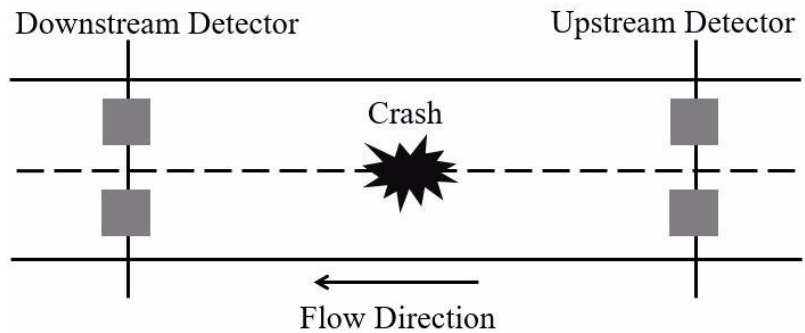

**Figure 1.** Detector location with respect to crash location

The 'normal' traffic data refers to data related to speed, flow, or occupancy at any location along the expressway, where no crashes occurred within a three-hour window before or after the recorded time. To create this dataset, information was collected for each detector every month. The data spanned a full 24-hour period and included at least one day from each day of the week, such as Sunday, Monday, Tuesday, and so forth, for every month throughout the study duration. In order to minimize bias, rather than selecting just one sample, four separate 'normal' datasets were formed. Each of these datasets comprised 1,000 randomly chosen samples from the complete 'normal' traffic dataset. **Table 2** provides an overview of the data, including speed, flow, and occupancy, utilized for both normal and pre-crash traffic conditions.

*3.3 Data pre-processing for Kerner's three-phase traffic flow theory*

According to existing literature [9], detector-based methods typically focus on road sections devoid of intersections or on/off ramps. In this study, the research team initially selected road sections without on/off ramps at random from both directions of the Shibuya 3 route within the Tokyo Metropolitan Expressway. Within these selected sections, a total of 148 crashes were recorded. Each of these chosen sections was equipped with a pair of detectors at both ends. To construct the pre-crash dataset, one hour of traffic flow data was extracted from the detectors corresponding to the crash time. Based on the observations, a minimum of one hour is necessary for a broad moving traffic jam to develop and subsequently transition into synchronized flow. For the normal traffic dataset, we randomly selected one-hour samples from these chosen sections, ensuring that no crashes had occurred in the corresponding direction of the expressway for at least one hour before and after the selected time.

**Table 2.** Descriptive statistics of data prepared for classical two-phase traffic flow theories

| Variables | Statistics | Normal traffic condition | | | | Pre-crash traffic condition (UD/DD) | | | | | | | | |
|---|---|---|---|---|---|---|---|---|---|---|---|---|---|---|
| | | ND1 | ND2 | ND3 | ND4 | T=1 | T=2 | T=3 | T=4 | T=5 | T=6 | T=7 | T=8 | T=9 |
| Occupancy (%) | Minimum | 0.30 | 0.40 | 0.40 | 0.30 | 0.00/ 0.00 | 0.00/ 0.00 | 0.00/ 0.00 | 0.00/ 0.00 | 0.00/ 0.00 | 0.00/ 0.00 | 0.00/ 0.00 | 0.00/ 0.00 | 0.00/ 0.00 |
| | Maximum | 69.80 | 39.35 | 48.25 | 66.45 | 68.65/ 90.50 | 84.15/ 90.15 | 100.00/ 77.40 | 71.65/ 84.15 | 82.40/ 84.15 | 98.65/ 94.35 | 71.30/ 77.55 | 99.05/ 93.25 | 93.00/ 77.00 |
| | Mean | 11.39 | 8.81 | 8.53 | 9.95 | 17.60/ 20.29 | 17.99/ 20.23 | 17.55/ 20.20 | 17.51/ 20.13 | 17.60/ 20.13 | 18.27/ 19.75 | 18.10/ 19.01 | 18.53/ 18.47 | 18.12/ 18.22 |
| | Standard Deviation | 10.44 | 7.21 | 6.95 | 9.31 | 14.65/ 16.63 | 15.54/ 16.36 | 15.09/ 16.01 | 15.02/ 16.09 | 14.65/ 16.09 | 15.17/ 15.82 | 14.33/ 15.03 | 15.39/ 14.94 | 14.56/ 14.57 |
| Flow (veh/min) | Minimum | 0.00 | 1.00 | 1.00 | 1.00 | 0.00/ 0.00 | 0.00/ 0.00 | 0.00/ 0.00 | 0.00/ 0.00 | 0.00/ 0.00 | 0.00/ 0.00 | 0.00/ 0.00 | 0.00/ 0.00 | 0.00/ 0.00 |
| | Maximum | 69.00 | 70.00 | 67.00 | 76.00 | 63.00/ 61.00 | 61.00/ 63.00 | 61.00/ 66.00 | 62.00/ 64.00 | 69.00/ 64.00 | 66.00/ 67.00 | 64.00/ 68.00 | 68.00/ 67.00 | 65.00/ 62.00 |
| | Mean | 32.17 | 29.39 | 30.93 | 30.50 | 26.95/ 26.94 | 26.76/ 27.12 | 27.44/ 27.35 | 28.14/ 28.67 | 29.06/ 28.67 | 29.40/ 29.82 | 29.63/ 29.96 | 29.99/ 30.62 | 30.68/ 30.73 |
| | Standard Deviation | 14.76 | 14.66 | 14.38 | 15.04 | 12.09/ 12.19 | 12.15/ 12.17 | 12.32/ 11.72 | 13.05/ 12.73 | 13.08/ 12.73 | 13.44/ 13.23 | 13.35/ 13.21 | 13.92/ 13.18 | 13.73/ 13.14 |
| Speed (km/hr) | Minimum | 0.00 | 10.20 | 0.00 | 4.55 | 0.00/ 0.00 | 0.00/ 0.00 | 0.00/ 0.00 | 0.00/ 0.00 | 0.00/ 0.00 | 0.00/ 0.00 | 0.00/ 0.00 | 0.00/ 0.00 | 0.00/ 0.00 |
| | Maximum | 117.85 | 110.00 | 115.40 | 117.85 | 94.45/ 102.00 | 96.90/ 100.00 | 100.00/ 100.00 | 107.60/ 92.30 | 110.10/ 92.30 | 93.60/ 100.00 | 95.50/ 117.85 | 101.55/ 98.10 | 105.90/ 100.00 |
| | Mean | 64.68 | 66.91 | 67.98 | 65.48 | 42.11/ 39.38 | 42.26/ 39.10 | 43.42/ 39.89 | 43.39/ 40.44 | 43.58/ 40.44 | 42.84/ 41.72 | 42.43/ 42.72 | 42.74/ 43.65 | 43.41/ 44.62 |
| | Standard Deviation | 24.56 | 18.82 | 17.61 | 21.04 | 27.19/ 27.29 | 27.40/ 27.46 | 27.33/ 27.50 | 26.85/ 27.17 | 26.80/ 27.17 | 26.88/ 27.14 | 26.21/ 27.35 | 26.89/ 26.94 | 26.91/ 27.25 |

ND: Normal traffic dataset, UD: Upstream dataset, DD: downstream dataset

## 4 RESULT AND DISCUSSIONS

In context of two-phase theory, the analysis unfolded in several stages. Firstly, a comprehensive analysis was carried out on the four sets of normal traffic data. This involved generating scatter plots for speed-flow-occupancy, establishing linear transformations to represent the four classical relationships. In the next phase, the classical traffic flow relationships were extracted for the 18 pre-crash datasets. Then, hypothesis tests were applied to explore whether the 'pre-crash' relationships exhibited significant variations from the 'normal' relationships. Finally, the research findings were synthesized and systematically presented within the framework of two-phase traffic flow theory. For three-phase traffic flow, a comparative analysis was conducted to assess wide-moving jam velocities between detector-based and correlation-based methods.

### 4.1 Deriving relationships for normal traffic condition

**Figure 2** illustrates the matrix scatter plot developed using occupancy and speed data under normal traffic condition (ND1) for the four classical traffic flow models. **Table 3** provides the corresponding coefficient of determination ($R^2$), as well as the values for occupancy factor at jam density ($K_j$), free flow speed ($V_f$), occupancy factor for density at maximum flow ($K_0$), and speed at maximum flow ($V_0$) values for all four datasets.

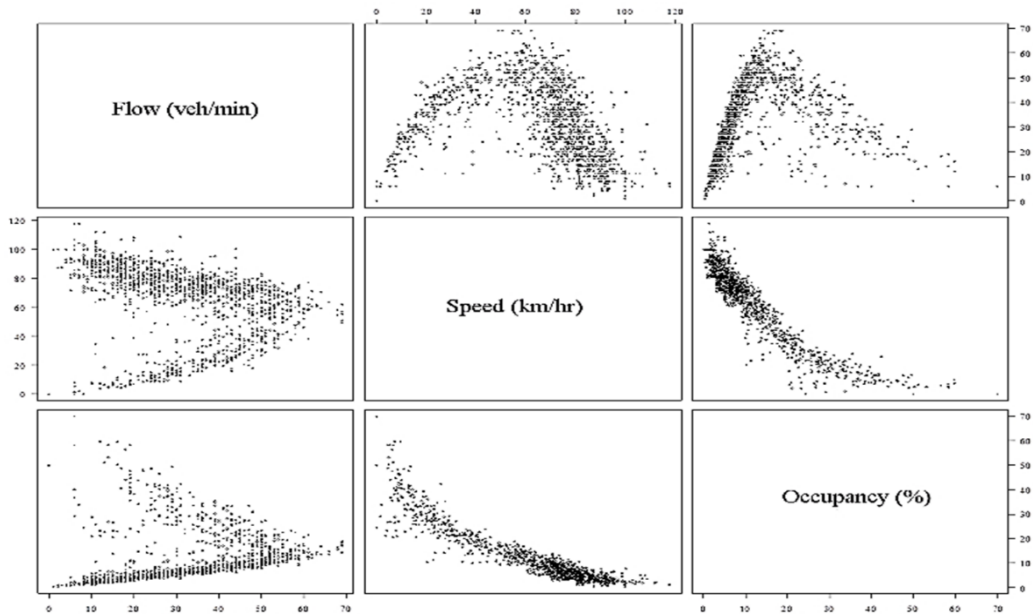

**Figure 2.** Matrix of scatter plots for normal traffic flow data (ND1)

**Table 3.** Estimated model parameters value for normal traffic conditions

| Dataset | Model | $R^2$ | $V_f$ | $V_0$ | $K_0$ | $K_j$ |
|---|---|---|---|---|---|---|
| ND1 | GS | 0.84 | 93.95 | | | 36.57 |
| | GR | 0.80 | | 32.34 | | 59.24 |
| | UW | 0.86 | 119.86 | | 15.13 | |
| | BS | 0.75 | 82.43 | | 17.67 | |
| ND2 | GS | 0.76 | 94.05 | | | 32.36 |
| | GR | 0.67 | | 34.42 | | 48.51 |
| | UW | 0.84 | 111.72 | | 14.91 | |
| | BS | 0.75 | 80.05 | | 17.00 | |

| Dataset | Model | $R^2$ | $V_f$ | $V_0$ | $K_0$ | $K_j$ |
|---------|-------|-------|-------|-------|-------|-------|
| ND3 | GS | 0.74 | 95.58 | | | 31.41 |
| | GR | 0.58 | | 30.25 | | 64.79 |
| | UW | 0.79 | 113.88 | | 15.46 | |
| | BS | 0.71 | 82.54 | | 16.45 | |
| ND4 | GS | 0.76 | 94.23 | | | 32.56 |
| | GR | 0.60 | | 32.19 | | 55.41 |
| | UW | 0.83 | 112.31 | | 15.86 | |
| | BS | 0.83 | 80.79 | | 17.02 | |

The relationships between speed-flow, speed-occupancy and flow-occupancy relationships in this study align with previous literature findings (**Figure 3**). As shown in **Table 3**, the values of $R^2$ were notably high for Greenshields, Underwood and Bell-shaped models, consistently explaining over 75% of the speed variation when estimating occupancy. For these models, the values of $R^2$ ranged from 0.74 to 0.84 for Greenshields, 0.79 to 0.86 for Underwood, and 0.71 to 0.83 for the Bell-shaped model, across four datasets. However, the $R^2$ values were lower for Greenberg model, ranging from 0.58 to 0.67, except for ND1, where it reached 0.8. It is important to note that expressways typically operate at high speeds with low traffic density, which could have influenced the Greenberg model's relatively poor fit, as it is known to perform less reliably under such traffic conditions. Despite the similarity of free flow speeds among the models within the datasets, e.g., ranging from 94 to 96 km/hr for the Greenshields model, there were variations among different models. For example, the Underwood model estimated a high free flow speed (120 km/hr in ND1), while the Bell-shaped model estimated a lower value (120 km/hr in ND2). In general, Greenshields models provided a free flow speed close to the average of these two values. Regarding the speed at maximum flow conditions ($V_0$), which is an output of the Greenberg model, it exhibited slight variations among the four datasets, ranging from 30 to 34 km/hr. The occupancy factors for maximum flow condition ($K_0$) remained relatively consistent for both Underwood (ranging from 14.91 to 15.86) and Bell-shaped models (ranging from 16.45 to 17.67). In contrast, the occupancy factors for jam density ($K_j$) showed substantial variations, with Greenshields ranging from 31.41 to 36.57 and Greenberg from 48.51 to 64.79, although the variations within the four Greenshields models were slight.

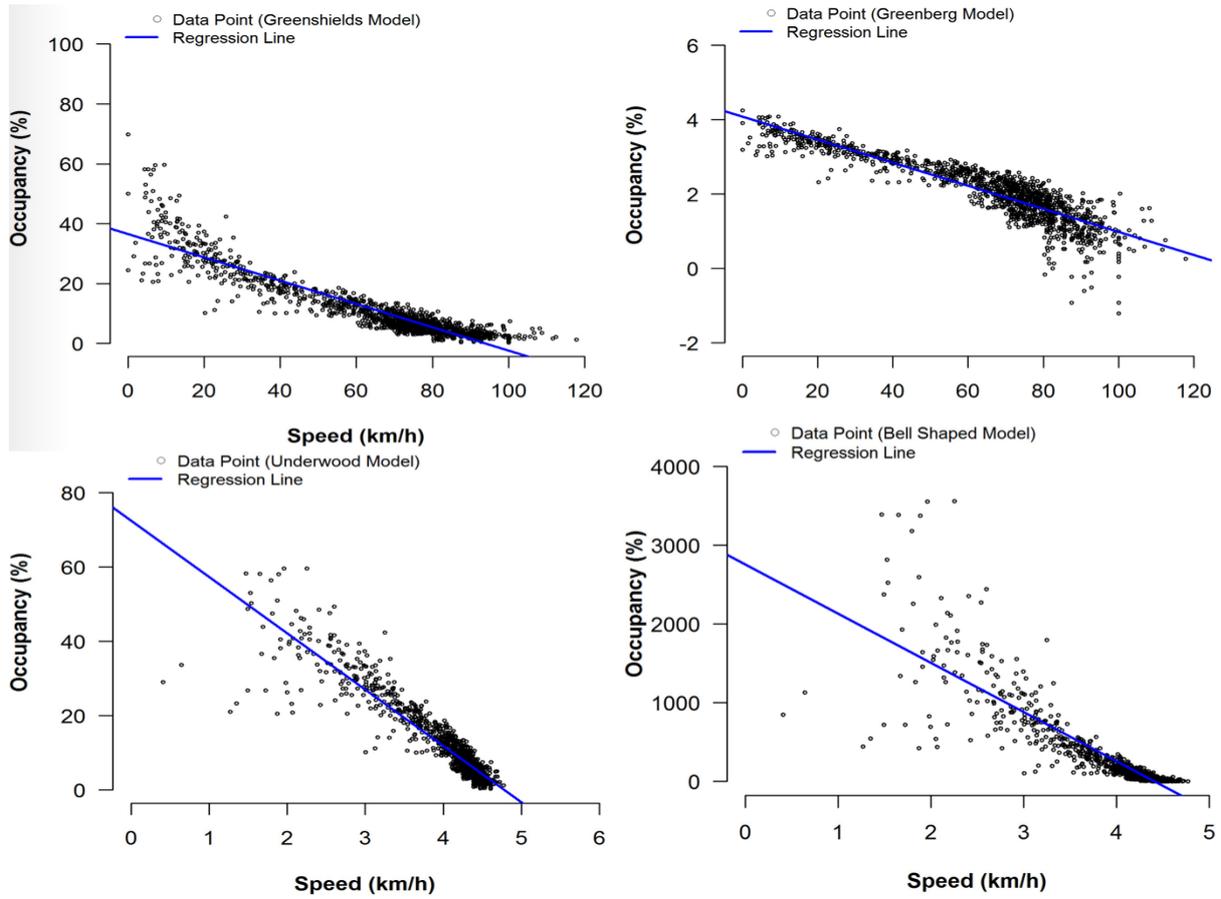

**Figure 3.** Regression lines in four models for Normal traffic data (ND1)

*4.2 Comparing slopes for linear transformation of normal traffic condition data*

    The analysis results in the preceding sub-section indicate the variation among the four datasets is relatively low when examined individually within the context of the four classical traffic flow theories, particularly for the Greenshields, Underwood, and Bell-shaped models. However, to assess whether there are variations between 'pre-crash' and 'normal' traffic conditions, it is crucial to confirm that the four datasets representing normal traffic conditions are indeed representative of the broader population of normal traffic. To accomplish this, hypothesis tests were conducted. Initially, none of the normal traffic condition datasets could be definitively considered as representative of the entire population. Therefore, all possible combinations were tested, with each test involving the selection of two datasets at a time. Hence, a total of $4 \times {}^4C_2 = 24$ combinations were examined, and the results are presented in **Table 4**, comparing the p-values with the level of significance. In the table, 'Y' indicates that the p-value is less than $\alpha$, signifying a significant difference, while 'X' suggests the vice-versa. Additionally, the notation ND1:ND2 indicates that the comparison was made between ND1 and ND2, and so forth. The outcome suggests that there were no significant differences among the four datasets while linearly fitting the Greenshields and Underwood models. However, the linear transformations of Greenberg models constructed with ND2 and ND3, as well as Bell-shaped models with ND3 and ND4, were found to be significantly different. This implies that these models may not have originated from the same population. The observed lower goodness-of-fit for Greenberg and Bell-shaped models for ND3 may further substantiate these findings.

**Table 4.** Hypothesis test for slope comparison among normal datasets

| Model | ND1:ND2 | ND1:ND3 | ND1:ND4 | ND2:ND3 | ND2:ND4 | ND3:ND4 |
|---|---|---|---|---|---|---|
| GS | X | X | X | X | X | X |
| GB | X | X | X | Y | X | X |
| UW | X | X | X | X | X | X |
| BS | X | X | X | X | X | Y |

GS = Greenshields, GB = Greenberg, UW = Underwood, BS = Bell-Shaped

*4.3 Pre-crash traffic condition at nearest upstream detector (UD)*

Data was collected from one minute to nine minutes before each crash, specifically from the nearest upstream detector, and nine distinct models were created for each of the four classical traffic flow theories. The results, including the coefficient of determination ($R^2$) values and model parameters, are presented in **Table 5**. The overall goodness of fit for all four models across all time intervals was notably lower compared to the four sets of normal traffic data. Except for the Underwood models, which performed the best under normal traffic conditions, the other three models consistently exhibited $R^2$ values below 0.7 in nearly all cases. No specific trend could be discerned by analyzing the variations in $R^2$ values over the nine-minute intervals for any of the models. Regarding the values of $V_f$, Greenshields and Bell-Shaped models consistently indicated lower values for pre-crash conditions, ranging from 86.18 to 89.07 km/hr for Greenshields and 67.36 to 73.03 km/hr for Bell-Shaped, compared to the normal traffic data, where they ranged from 93.95 to 95.58 km/hr for Greenshields and 80.05 to 82.54 km/hr for Bell-Shaped models. In contrast, the Underwood model estimated a higher $V_f$ value, ranging from 109.64 to 126.42 km/hr, while the normal traffic conditions suggested a range of 111.72 to 119.86 km/hr. The speed at maximum flow, denoted as $V_0$, indicated a higher speed (ranging from 35.32 to 37.12 km/hr) during pre-crash situations compared to normal traffic conditions (ranging from 30.25 to 34.42 km/hr). The $K_0$ values, obtained from the Underwood and Bell-Shaped models, ranged between 14.1 to 15.11 and 18.62 to 20.37, respectively, suggesting a higher level of traffic compactness during pre-crash traffic conditions compared to normal traffic conditions at maximum flow. Finally, the $K_j$ values exhibited a significant difference between the Greenshields and Greenberg models. The former (ranging from 34.41 to 36.25) suggested a more compact traffic condition at jam density, while the latter (ranging from 41.97 to 44.95) indicated less compactness when compared to normal traffic conditions (ranging from 31.41 to 36.57 for Greenshields and 48.51 to 64.79 for Greenberg). In summary, most of the crashes occurred during congested periods in the study area, and the free flow speed ($V_f$) was lower in pre-crash conditions. Greenshields, Underwood, and Bell-Shaped models suggested a maximum flow condition in which vehicles traveled more closely but at higher speeds than in normal traffic conditions, which could be associated with an elevated crash risk.

**Table 5.** Estimated model parameters for pre-crash traffic conditions (upstream detector)

| Model | Parameter | T=1 | T=2 | T=3 | T=4 | T=5 | T=6 | T=7 | T=8 | T=9 |
|---|---|---|---|---|---|---|---|---|---|---|
| GS | $V_f$ | 86.18 | 86.18 | 87.58 | 86.23 | 86.81 | 86.89 | 87.26 | 87.42 | 89.07 |
|  | $K_j$ | 34.42 | 35.72 | 34.81 | 35.24 | 35.35 | 36.04 | 35.23 | 36.25 | 35.34 |
|  | $R^2$ | 0.55 | 0.55 | 0.52 | 0.53 | 0.56 | 0.54 | 0.55 | 0.53 | 0.54 |
| GB | $V_0$ | 35.67 | 35.32 | 35.57 | 36.08 | 36.16 | 36.52 | 36.57 | 37.12 | 36.57 |
|  | $K_j$ | 41.97 | 43.50 | 43.39 | 42.06 | 43.72 | 44.04 | 43.53 | 43.60 | 44.95 |
|  | $R^2$ | 0.66 | 0.70 | 0.69 | 0.61 | 0.70 | 0.67 | 0.65 | 0.64 | 0.69 |
| UW | $V_f$ | 112.04 | 109.64 | 115.68 | 112.29 | 117.42 | 120.67 | 123.72 | 121.15 | 126.42 |
|  | $K_0$ | 14.63 | 15.11 | 14.55 | 14.97 | 14.51 | 14.37 | 14.13 | 14.57 | 14.10 |

| Model | Parameter | T=1 | T=2 | T=3 | T=4 | T=5 | T=6 | T=7 | T=8 | T=9 |
|---|---|---|---|---|---|---|---|---|---|---|
|  | $R^2$ | 0.76 | 0.78 | 0.71 | 0.75 | 0.74 | 0.73 | 0.71 | 0.73 | 0.70 |
| BS | $V_f$ | 69.97 | 67.36 | 72.72 | 68.62 | 70.61 | 71.16 | 73.03 | 67.31 | 71.92 |
|  | $K_0$ | 18.74 | 19.84 | 18.84 | 19.63 | 19.10 | 19.15 | 18.62 | 20.37 | 19.16 |
|  | $R^2$ | 0.63 | 0.62 | 0.50 | 0.61 | 0.59 | 0.59 | 0.59 | 0.55 | 0.54 |

### *4.4 Pre-crash traffic condition at nearest downstream detector (DD)*

Nine models were created using pre-crash data collected from the closest downstream detectors relative to the crash location, and their outcomes are shown in **Table 6**. Similar to the models constructed with upstream pre-crash data, these models also displayed a lower level of goodness-of-fit when compared to the datasets representing four normal traffic conditions. However, the $R^2$ values were slightly higher for all four classical models for the pre-crash data obtained from downstream detectors. The Underwood models demonstrated the best performance, with $R^2$ values ranging from 0.78 to 0.82. Notably, the Greenberg models also exhibited significant improvement, with their $R^2$ values varying from 0.66 to 0.78. Nevertheless, there was no discernible pattern in the performance variation among these models over time. In comparison to the models constructed using upstream data, these models indicated a lower free-flow speed ($V_f$), which was notably lower than that of normal traffic conditions for all four classical models. Specifically, the values for the Greenshields, Underwood, and Bell-Shaped models ranged from 82.26 to 88.38, 103.85 to 114.24, and 67.78 to 69.98 km/hr, respectively. The speed at maximum flow ($V_0$), as derived by the Greenberg model, was also lower (ranging from 32.57 to 36.46 km/hr) compared to the upstream detector data but still higher than the values for normal traffic conditions (ranging from 30.25 to 34.42 km/hr). In terms of the occupancy factors for density at maximum flow ($K_0$), the data from downstream detectors suggested a higher level of traffic compactness (ranging from 15.53 to 16.21 for Underwood and 19.76 to 20.38 for Bell-Shaped) in comparison to the upstream detectors for pre-crash traffic conditions. This trend continued for the occupancy factors at jam density ($K_j$), with the Greenshields model showing a higher level of compactness. However, the corresponding values for Greenberg models were higher than those from the upstream dataset but still considerably lower than those from normal traffic conditions.

**Table 6.** Estimated model parameters for pre-crash traffic conditions (downstream detector)

| Model | Parameter | T=1 | T=2 | T=3 | T=4 | T=5 | T=6 | T=7 | T=8 | T=9 |
|---|---|---|---|---|---|---|---|---|---|---|
| GS | $V_f$ | 82.26 | 83.03 | 83.87 | 85.26 | 84.39 | 85.78 | 87.28 | 87.57 | 88.38 |
|  | $K_j$ | 38.92 | 38.24 | 38.52 | 37.45 | 38.65 | 38.45 | 37.23 | 36.82 | 36.80 |
|  | $R^2$ | 0.60 | 0.60 | 0.63 | 0.59 | 0.60 | 0.59 | 0.60 | 0.58 | 0.61 |
| GB | $V_0$ | 32.64 | 33.39 | 32.57 | 34.19 | 33.69 | 34.38 | 34.81 | 35.41 | 36.46 |
|  | $K_j$ | 47.99 | 46.86 | 49.17 | 45.91 | 48.44 | 48.86 | 48.02 | 47.65 | 45.73 |
|  | $R^2$ | 0.74 | 0.74 | 0.78 | 0.70 | 0.73 | 0.74 | 0.76 | 0.75 | 0.67 |
| UW | $V_f$ | 103.85 | 107.26 | 107.06 | 109.13 | 111.02 | 113.38 | 111.70 | 112.11 | 114.24 |
|  | $K_0$ | 16.21 | 15.53 | 15.97 | 15.53 | 15.78 | 15.76 | 15.88 | 15.87 | 15.67 |
|  | $R^2$ | 0.78 | 0.81 | 0.79 | 0.78 | 0.79 | 0.77 | 0.79 | 0.79 | 0.82 |
| BS | $V_f$ | 65.64 | 64.78 | 68.90 | 68.30 | 68.24 | 69.98 | 69.96 | 68.32 | 69.62 |
|  | $K_0$ | 20.71 | 20.48 | 19.97 | 19.76 | 20.37 | 20.25 | 19.94 | 20.38 | 20.04 |
|  | $R^2$ | 0.59 | 0.68 | 0.64 | 0.66 | 0.65 | 0.60 | 0.66 | 0.64 | 0.71 |

While the trends in both the pre-crash datasets (upstream and downstream) shared similarities, the downstream data provided a clearer distinction with normal traffic conditions. This was evident through more substantial differences in key variables like free-flow speed ($V_f$), speed at maximum flow ($V_0$), density at maximum flow ($K_0$), and occupancy factors at jam density ($K_j$). As a result, it can be inferred that the nearest downstream detectors, in relation to the crash locations, offer a more effective explanation of traffic conditions in the context of crashes. Among the four classical models, the Underwood models, both for upstream and downstream data, displayed superior goodness-of-fit. Furthermore, when comparing the results from both upstream and downstream detectors, it becomes evident that several crashes in the study area occurred at points where fast-moving traffic from upstream encountered relatively denser and slower-moving traffic downstream.

*4.5 Comparing slopes of normal and pre-crash data of estimated parameters*

Hypothesis tests were conducted to examine whether the disparities in the slopes of models derived from pre-crash datasets and normal datasets are statistically significant. Given that **Table 4** suggests that only ND1 can be regarded as originating from the same population, while comparing with other normal traffic condition datasets. Subsequently, each of models generated from ND1 was compared with the corresponding models for both upstream and downstream datasets across all nine pre-crash time segments. This resulted in a total of 72 slope comparisons (4 classical models x 2 upstream and downstream detectors x 9 time segments), as presented in **Table 7**. In the table, the outcomes are denoted as either the null hypothesis being rejected (Y) or not rejected (X). Detector positions (upstream and downstream) and the time segment are represented as XD_t, where X stands for U (upstream) or D (downstream), and t ranges from 1 to 9. The findings from Table 7 indicate that, except for the linear transformation of the Underwood model using data from the immediate upstream detectors just two minutes before the crashes, all other models for all the other time segments were found to be significantly different from the traffic conditions that prevailed during periods without crash hazards in the study area.

**Table 7.** Hypothesis test for slope comparison (normal vs. pre-crash)

| Model | UD_1 | UD_2 | UD_3 | UD_4 | UD_5 | UD_6 | UD_7 | UD_8 | UD_9 |
|---|---|---|---|---|---|---|---|---|---|
| GS | Y | Y | Y | Y | Y | Y | Y | Y | Y |
| GB | Y | Y | Y | Y | Y | Y | Y | Y | Y |
| UW | Y | X | Y | Y | Y | Y | Y | Y | Y |
| BS | Y | Y | Y | Y | Y | Y | Y | Y | Y |
|  | DD_1 | DD_2 | DD_3 | DD_4 | DD_5 | DD_6 | DD_7 | DD_8 | DD_9 |
| GS | Y | Y | Y | Y | Y | Y | Y | Y | Y |
| GB | Y | Y | Y | Y | Y | Y | Y | Y | Y |
| UW | Y | Y | Y | Y | Y | Y | Y | Y | Y |
| BS | Y | Y | Y | Y | Y | Y | Y | Y | Y |

*4.6 Comparison of pre-crash and normal traffic for Kerner's three-phase traffic flow theory*

For each set of pre-crash data, graphs were generated illustrating the relationship between flow and time, as well as the corresponding speed and time graphs. Consecutive numerical time stamps were assigned to each time interval within the selected duration. These graphical representations serve the purpose of identifying the presence of a wide-moving traffic jam phase during a specific timeframe. Such a wide-moving jam phase is characterized by a simultaneous decrease in both traffic flow and speed. An example is provided in **Figure 4**, where a time duration from 15:40 to 16:40 is chosen with 1-minute time intervals. As a result, the first interval, labeled '15:40 -15:41', is assigned a time stamp of 1, and the final interval,

'16:39-16:40', is designated with a time stamp of 60. Furthermore, it is evident from the figure that after the 40th minute, a substantial drop in both flow and speed occurred simultaneously, indicating the presence of a wide-moving traffic jam.

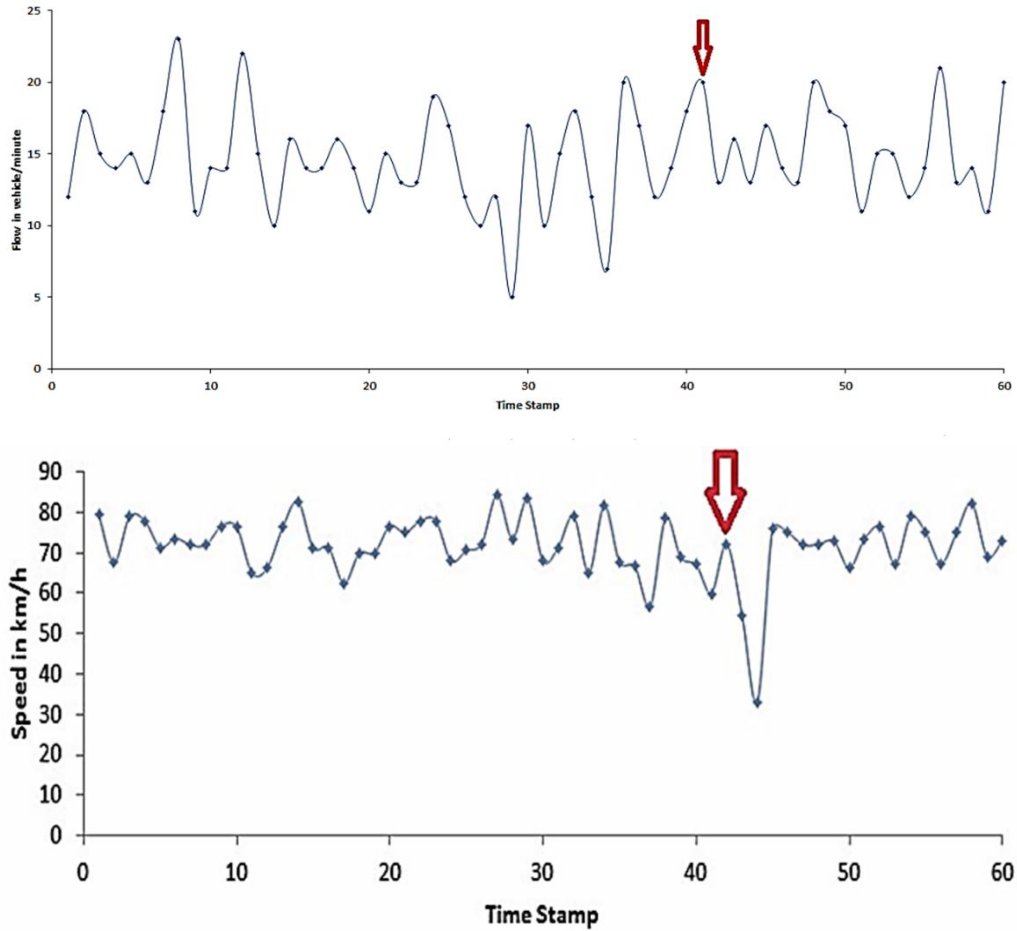

**Figure 4.** (a) flow versus time graph for one hour interval, (b) its corresponding speed versus time graph.

Following the identification of the wide-moving traffic jam, similar to the approach used in detector-based methods, a time span of 25 minutes was selected from within the duration of the jam. Subsequently, the flow correlation between the upstream and downstream detectors, along with the corresponding wave velocity, was computed for time lags ranging from 1 to 15 minutes. Using these results, a graph depicting correlation values against velocity, as exemplified in **Figure 5**, was generated. In this figure, the velocity corresponding to the highest correlation value is identified as the downstream jam velocity. As demonstrated in **Figure 5**, the maximum correlation value, reaching 0.55, was observed at a velocity of 13.89 km/h.

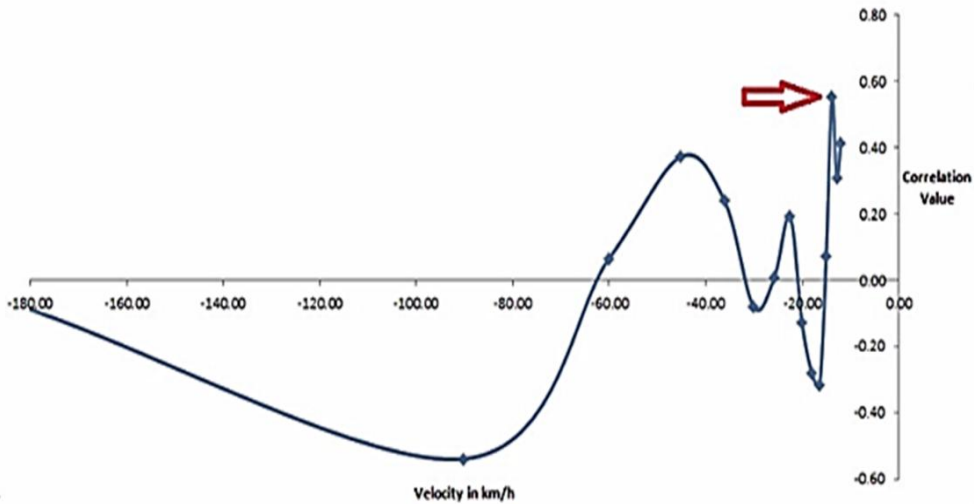

**Figure 5.** Correlation versus velocity diagram

The same computation was conducted for all 148 crash samples, encompassing their corresponding pre-crash and normal traffic data. **Table 8** provides an overview of the minimum, maximum, average downstream jam velocities, and their respective standard deviations for different methods and conditions. Remarkably, all values fall within the range of -10 to -18 kph, in line with the principles of the two-phase traffic flow theory. This consistency extended to both pre-crash and normal traffic conditions, reinforcing their validity within the context of three-phase traffic flow theory as well. The results underscore a noteworthy pattern where the pre-crash traffic exhibited a notably higher wide-moving jam velocity that propagates from downstream to upstream through the traffic compared to the conditions observed in normal traffic.

**Table 8.** Statistics of the wide moving jam velocities for pre-crash and normal conditions

|  | Detector Based Method | | Correlation Based Method | |
|---|---|---|---|---|
|  | **Pre-crash** | **Normal** | **Pre-crash** | **Normal** |
| **Maximum** | -18.60 | -13.89 | -18.06 | -12.04 |
| **Minimum** | -11.10 | -10.62 | -11.93 | -11.29 |
| **Mean** | -14.38 | -12.37 | -14.32 | -11.67 |
| **Standard Deviation** | 2.27 | 1.34 | 2.33 | 0.31 |

## 5 CONCLUSIONS

This study employs loop detector data from the Shibuya 3 and Shinjuku 4 routes situated within the Tokyo Metropolitan Expressway to distinguish variations between pre-crash and normal traffic flow conditions within the context of classical two-phase and Kerner's three-phase traffic flow theories. The results indicate that all four classical models of the two-phase traffic flow theory demonstrate a strong fit for normal traffic data. Except for the Greenberg models, the estimated $R^2$ values for the Greenshields, Underwood, and Bell-Shaped models exceeded 0.75 for all four sets of normal traffic data. Notably, the data from the nearest downstream detectors to the crash locations provided a better fit for the pre-crash data in comparison to the upstream detectors. For pre-crash conditions, data was collected from both the nearest upstream and downstream detectors every minute, up to nine minutes before the crash. A marked decrease in goodness-of-fit was observed in comparison to normal traffic conditions. However, similar to models using normal traffic data, all models with pre-crash data remained statistically significant. Among these, the Underwood and Greenberg models exhibited superior goodness-of-fit for both upstream and downstream pre-crash data, while Greenshields performed less effectively with the same datasets. The values for free-flow speed, speed at maximum flow, and parameters related to jam density and density at maximum flow indicated that the flow of vehicles was more compacted during pre-crash conditions compared to normal conditions. Furthermore, during maximum flow conditions, the pre-crash traffic moved at a higher speed than normal traffic. This trend was consistent across all four classical models and suggests traffic conditions that may demand a higher level of cognitive ability from the drivers. Additionally, a recurring pattern was observed where a faster-moving but compact upstream met compact yet relatively slower downstream traffic just before a crash. Both upstream and downstream pre-crash conditions resulted in the rejection of the null hypothesis that there was no difference between pre-crash and normal conditions with a 95% confidence interval. Similar results were obtained from the three-phase traffic flow theory, as both pre-crash and normal traffic data met the criteria for wide-moving jams in both calculation methods. Consistent with the findings from the two-phase theory, it was observed that wide-moving jams moved faster from downstream to upstream during pre-crash conditions compared to normal traffic conditions. Additionally, pre-crash data exhibited greater standard deviation in the calculated wide-moving jam velocities compared to normal traffic conditions. These findings can prove valuable for the development of models for predicting crashes and for estimating traffic volumes on freeways across various timeframes, taking into account both free-flow and congested traffic scenarios.

This study is subject to certain limitations. It exclusively employed two out of four potential methods for its investigation, namely the detector-based method and the correlation method. The remaining two methods, namely the graphical method and the flow density method, were not utilized. The graphical method necessitates the use of graphs generated by the ASDA/FOTO software. Future research could expand on these limitations by incorporating these additional methods to gain a more comprehensive understanding of the associated differences.

**ACKNOWLEDGEMENT**
The authors would like to thank Tokyo Metropolitan Expressway Company Limited for providing the traffic data to conduct the study. A part of this research has been published in Proceedings of the 12th International Conference of Eastern Asia Society for Transportation Studies, At Ho Chi Minh City, Vietnam.

**DECLARATION of CONFLICTING INTERESTS**
The author(s) declared no potential conflicts of interest with respect to the research, authorship, and/or publication of this article.

**STUDY CONTRIBUTIONS**
The authors confirm contribution to the paper as follows: study conception and design: M.M. Hossain; Data preparation: M.M. Hossain; analysis and interpretation of results: M.M. Hossain; draft manuscript preparation: M.M. Hossain. All authors reviewed the results and approved the final version of the manuscript.